\documentclass[twocolumn,showpacs,preprintnumbers,amsmath,amssymb,prl]{revtex4}
\usepackage{amsfonts}
\usepackage[english]{babel}
\usepackage[T1]{fontenc}
\usepackage{times}
\usepackage{mathrsfs}
\usepackage{graphicx}
\usepackage{dcolumn}
\usepackage{bm}
\usepackage{wasysym}
\usepackage[colorlinks,bookmarks=true,citecolor=blue,linkcolor=red,urlcolor=blue]{hyperref}
\usepackage[tight, FIGTOPCAP, hang, raggedright, nooneline]{subfigure}
\usepackage{hyperref}
\usepackage[makeroom]{cancel}

\begin{document}

	\title{Excitons in topological Kondo insulators -- \\theory of thermodynamic and transport anomalies in SmB$_6$}
		\author{Johannes \surname{Knolle*}}
		\affiliation{T.C.M. Group, Cavendish Laboratory, J.J. Thomson Avenue, Cambridge CB3 0HE, United Kingdom}
		\author{Nigel R. \surname{Cooper}}
		\affiliation{T.C.M. Group, Cavendish Laboratory, J.J. Thomson Avenue, Cambridge CB3 0HE, United Kingdom}
		
		\date{\today}

		\begin{abstract}
Kondo insulating materials lie outside the usual dichotomy of weakly versus correlated -- band versus Mott -- insulators. They are metallic at high temperatures but resemble band insulators at low temperatures because of the opening of an interaction induced band gap. The first discovered Kondo insulator (KI) SmB$_6$ has been predicted to form a topological KI (TKI) which mimics a topological insulator at low temperatures. However, since its discovery thermodynamic and transport anomalies have been  observed that have defied a theoretical explanation. Enigmatic signatures of collective modes inside the charge gap are seen in specific heat, thermal transport and quantum oscillation experiments in strong magnetic fields. Here, we show that TKIs are susceptible to the formation of excitons and magneto-excitons. These charge neutral composite particles can account for long-standing anomalies in SmB$_6$ which is crucial for the identification of bulk topological signatures. 
		\end{abstract}
		
\maketitle

One of the biggest successes of quantum mechanics is the explanation of the distinction between metals and insulators. Traditionally, there are two different regimes: First, in weakly interacting systems  insulating behaviour arises from complete filling of Bloch bands with a gap to unoccupied states, whereas metals have partially filled bands giving 
a manifold of gapless excitations -- the Fermi surface (FS). Second, in strongly interacting systems repulsion forbids hopping of electrons leading to Mott insulators. However, a third possibility exists -- so called Kondo insulators (KI) -- where strong interactions between itinerant electrons and localized spins lead to a heavy band insulator at low temperatures~\cite{Hewson1993}. 
\begin{figure}[tb]
\centering
\includegraphics[width=0.7\linewidth]{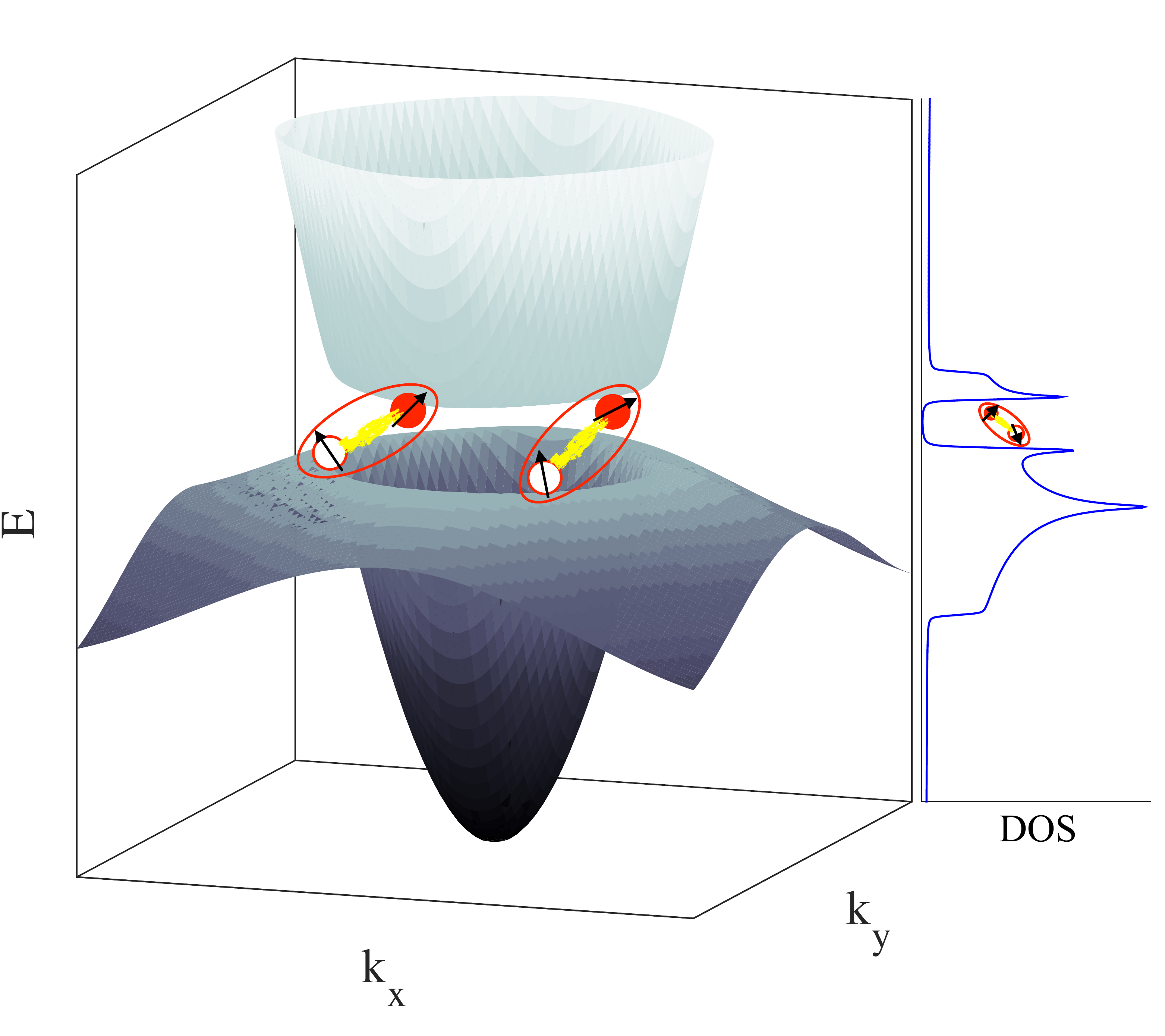}
\caption{{\bf Excitons in topological Kondo insulators.} At low temperatures strong correlations lead to the formation of fermionic quasiparticles which have the schematically shown band structure (for $k_z=0$) of a broad-brimmed Mexican hat. Due to the large DOS at the band edges it is very susceptible to the formation of excitons.}
\label{Fig1}
\end{figure}  
The material SmB$_6$ was the first KI discovered almost half a century ago \cite{Geballe1969}. More recently, it has been predicted that SmB$_6$ should be a topological KI (TKI) \cite{Dzero2010} which is an interaction-induced heavy 3D topological insulator~\cite{FuKane2007}.  Normally the charge gap  in an insulator also determines its thermodynamic and bulk transport properties which are expected to then follow an Arrhenius type activated temperature dependence. However, the tentative TKI SmB$_6$ strongly deviates from that picture and exhibits unusual thermodynamic anomalies, for example a low temperature specific heat contribution reminiscent of a metal~\cite{Nickerson1971,Phelan2014}. These have not found a generally accepted explanation for decades and more recent experiments motivated from the TKI proposal have added even bigger puzzles. The observation of bulk quantum oscillations (QO) \cite{Tan2015}, normally a synonym for a FS, and residual thermal transport inside the insulating low temperature regime~\cite{Suchitra2016} challenge our canonical understanding of metals and insulators.

Here, we show that due to the special Kondo-origin of the TKI, the system is very susceptible to the formation of {\it excitons} or {\it magneto-excitons} (MExc) in an  applied magnetic field $B$. The strong Coulomb repulsion of the localised Sm f-levels has two effects: First, it gives rise to a heavy insulating state with a peculiar broad-brimmed Mexican hat-like band structure, see Fig.\ref{Fig1}, which provides the necessary large density of states (DOS) from the band extrema, see right panel. Second, it provides the interaction which binds the particle-hole pairs. As our main result we establish that these composite quasiparticles inside the insulating gap and without charge degrees of freedom provide a natural  explanation of long standing anomalies in SmB$_6$.

\begin{figure}
\centering
\includegraphics[width=1.0\linewidth]{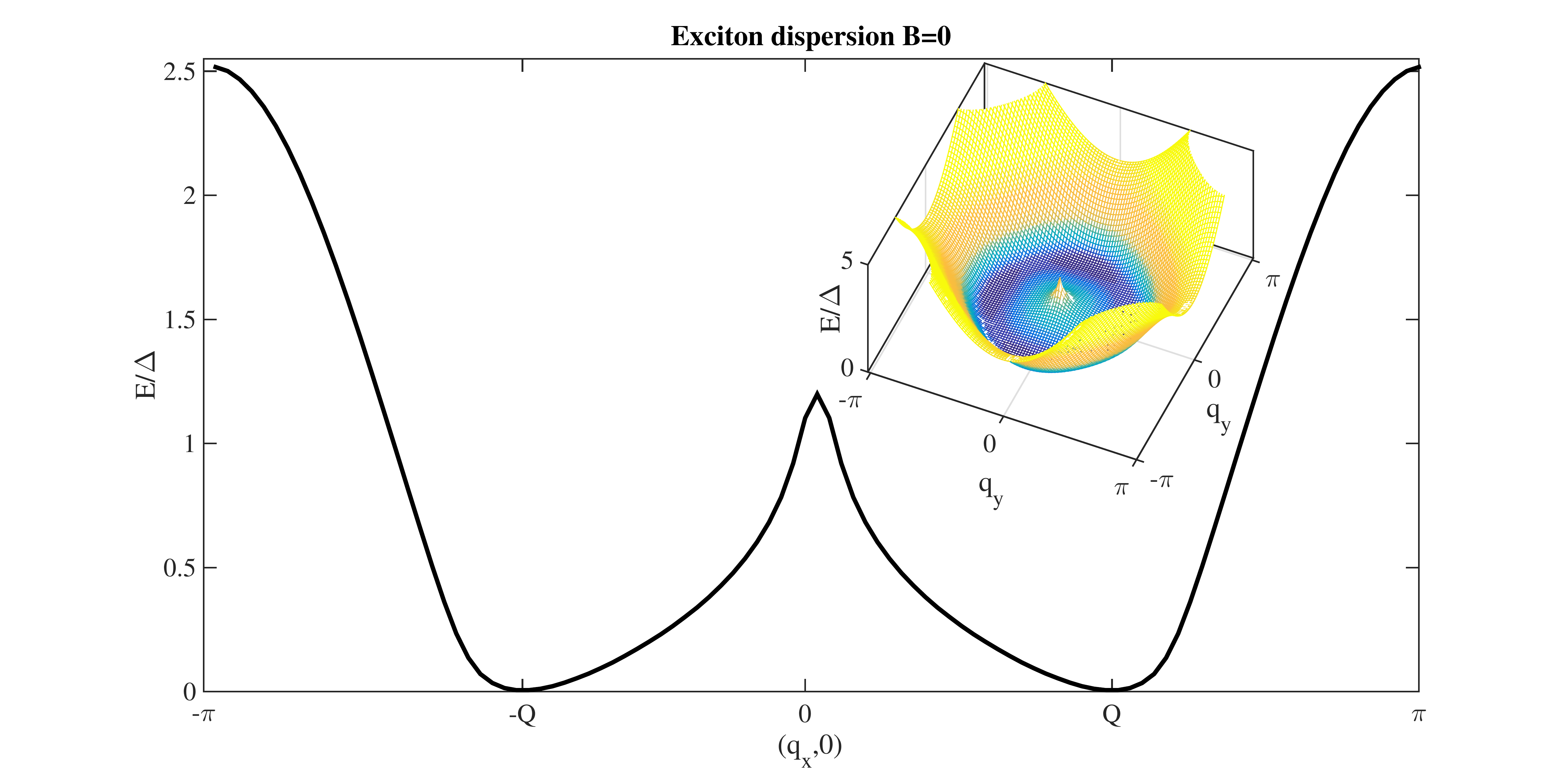}
\caption{{\bf Exciton dispersion.} Shown for $B=0$ as a function of $q_x$ with $q_y=0$. The full exciton dispersion of the 2D TKI is plotted in the inset. Because of the special band structure the dispersion minima are always at a nonzero momenta but their precise energy depends on the interaction strength. We have used parameters $W=\mu=-3.2$, $\alpha=0.15$, $\gamma=0.1$, $U=2.75$ (in units of hopping $t$), which give a dispersion with a small gap which can best explain experiments on SmB$_6$. 
}
\label{Fig1b}
\end{figure}  
    
{\it The model.}
We focus on a minimal model of a TKI which captures the essential physics of SmB$_6$ and takes the form of a periodic Anderson lattice model \cite{Dzero2010,Alexandrov2015}.  
\begin{eqnarray}
\label{Hamiltonian}
 H & = &  \sum_{\mathbf{k}, \alpha \beta} 
 \begin{pmatrix}
 d_{\mathbf{k},\alpha}^{\dagger} &  f_{\mathbf{k},\alpha}^{\dagger}\\
 \end{pmatrix}
\begin{pmatrix}
 \epsilon^d_{\mathbf{k}} & \frac{\gamma}{2} \vec s_{\mathbf{k}} \vec \sigma_{\alpha ,\beta} \\
 \frac{\gamma}{2} \vec s_{\mathbf{k}} \vec \sigma_{\alpha ,\beta}  &  \epsilon^f_{\mathbf{k}} \end{pmatrix}
\begin{pmatrix}
 d_{\mathbf{k},\beta} \\  f_{\mathbf{k},\beta}
\end{pmatrix}  \nonumber \\ 
& &   + U \sum_{i} f^{\dagger}_{\mathbf{r}_i,\uparrow}f_{\mathbf{r}_i,\uparrow}f^{\dagger}_{\mathbf{r}_i,\downarrow}f_{\mathbf{r}_i,\downarrow}
\end{eqnarray}
The broad Sm d-band has a dispersion $ \epsilon^d_{\mathbf{k}}=-2t\sum_{\mathbf{k},\eta=x,y,z} \cos k_{\eta}$. The almost flat inverted f-band is $ \epsilon^f_{\mathbf{k}}= -\alpha \epsilon^d_{\mathbf{k}}+\lambda$ (with $\alpha\ll 1$) with $\lambda=(1+\alpha)W$ and incorporates the strong on-site repulsion $U$. 
The d- and f-level have different angular momenta and the latter is a superposition of different spin components due to strong spin orbit coupling. Hence, the hybridisation in a TKI is spin and momentum-dependent (odd-parity), $\vec s_{\mathbf{k}}=\left ( \sin{k_x},\sin{k_y},\sin{k_z}\right )^{\rm T} $.

To account for the strongly correlated nature we treat the self energy of the f-electrons as momentum independent such that the main effect of the local  repulsion is a renormalisation of the f-bands and the hybridisation~\cite{Read1984}. The low energy excitations are described by a new non-interacting Hamiltonian identical to the first term in Eq.\ref{Hamiltonian} but with new parameters $\tilde t$ and $\tilde \gamma$~\cite{Ikeda1996} (we drop the tilde in the remainder). They are in general complicated functions of $U$ and temperature~\cite{Legner2014}. Here, we use effective parameters in agreement with the experimentally known dispersion around the three X-points of SmB$_6$~\cite{Lu2013,Neupane2013,Frantzeskakis2013}.

The Hamiltonian at low temperatures yields two two-fold degenerate energies 
$E^{\nu}_{\pm}(\mathbf{k}) = \frac{1}{2} \left[  \epsilon^d_{\mathbf{k}}+  \epsilon^f_{\mathbf{k}} \right] \pm d_{\mathbf{k}}$
with $d_{\mathbf{k}}=\frac{1}{2} \sqrt{\left( \epsilon^d_{\mathbf{k}}-  \epsilon^f_{\mathbf{k}} \right)^2+|\gamma s_{\mathbf{k}}|^2}$. A schematic form of the band structure is shown in Fig.\ref{Fig1}. In the TKI state the lower bands are completely filled and the broad-brimmed Mexican hat-like dispersion gives a large DOS near the gap. 
For simplicity, we concentrate in the following on an effective two-dimensional model but we have checked that the inclusion of the third dimension does not lead to any qualitative changes of our findings.   
Setting $k_z=0$ the Hamiltonian decouples into two independent blocks, labeled by $\nu=+1(-1)$, for the $d_{\uparrow}^{\dagger}f_{\downarrow}$ and $d_{\downarrow}^{\dagger}f_\uparrow$ states
diagonalised by 
\begin{eqnarray}
\label{EV}
c_{\mathbf{k},\nu} & = & \cos \frac{\beta_{\mathbf{k}}}{2} d_{\mathbf{k},\nu}  +  \sin \frac{\beta_{\mathbf{k}}}{2} e^{\text{i} \nu \theta_{\mathbf{k}}} f_{\mathbf{k},\bar \nu}  \\ \nonumber
v_{\mathbf{k},\nu} & = & -\sin \frac{\beta_{\mathbf{k}}}{2} e^{-\text{i} \nu \theta_{\mathbf{k}}}  d_{\mathbf{k},\nu}  +  \cos \frac{\beta_{\mathbf{k}}}{2} f_{\mathbf{k},\bar \nu}  
\end{eqnarray} 
with the angles given by $\cos \beta_{\mathbf{k}}=\frac{1}{2 d_{\mathbf{k}}} \left[  \epsilon^d_{\mathbf{k}}-  \epsilon^f_{\mathbf{k}} \right]$ and $\sin \beta_{\mathbf{k}} e^{-\text{i} \theta_{\mathbf{k}}} = \frac{\gamma}{2 d_{\mathbf{k}}}  \left (\sin{k_x}- \text{i}\sin{k_y} \right )$.

{\it Excitons.} To investigate the formation of bound excitons, we include the strong f-level interaction on top of these bands. We project the Hubbard term of Eq.\ref{Hamiltonian} onto the renormalised TKI bands and concentrate only on those terms which lead to exciton binding 
\begin{widetext}
\begin{eqnarray}
\label{HamiltonianValenceConduction}
H_{\text{int}}& = & -U \sum_{\mathbf{k},\mathbf{k'},\mathbf{q}} \left[ \phi(\mathbf{k+q,-q}) \phi^*(\mathbf{k'+q,-q}) c^{\dagger}_{\mathbf{k+q},-} v_{\mathbf{k},+} v^{\dagger}_{\mathbf{k'},+} c_{\mathbf{k'+q},-} +  \phi(\mathbf{k,q}) \phi^*(\mathbf{k',q}) v^{\dagger}_{\mathbf{k+q},-} c_{\mathbf{k},+} c^{\dagger}_{\mathbf{k'},+} v_{\mathbf{k'+q},-} \right]
\end{eqnarray}
\end{widetext}
with $ \phi(\mathbf{k,q})= \sin \frac{\beta_{\mathbf{k}}}{2} \cos \frac{\beta_{\mathbf{k+q}}}{2} e^{-\text{i} \theta_{\mathbf{k}}}$. Hence, our system is described by the Hamiltonian $H=H_0+H_{\text{int}}$ with $H_0 = \sum_{\mathbf{k},\nu} \left[ E_{+}^{\nu}(\mathbf{k}) c_{\mathbf{k},\nu}^{\dagger} c_{\mathbf{k},\nu} +  E_{-}^{\nu}(\mathbf{k}) v_{\mathbf{k},\nu}^{\dagger} v_{\mathbf{k},\nu} \right] $ and the 'valence' and 'conduction' bands $E^{\nu}_-$ and $E^{\nu}_+$ separated by a gap $\Delta=\text{min}(E^{\nu}_+)-\text{max}(E^{\nu}_-)$.
The interaction only binds electron-hole pairs of opposite $\nu$. For example, (-- +)-excitons [similarly for the time reversed partner (+ --)] are created by the  operator
\begin{eqnarray}
\label{ExcitonOperator}
S^{\dagger}_{\mathbf{q}}= \sum_{\mathbf{k}} \varphi_{\mathbf{q}}(\mathbf{k}) c^{\dagger}_{\mathbf{q},-} v_{\mathbf{k+q},+}
\end{eqnarray} 
which are directly related to spin flip excitations. 
We calculate their dispersion $E(\mathbf{q})$ from the Bethe-Salpeter equation
\begin{eqnarray}
\label{ExcitonEqOfMotion}
\left[ H_0+H_{\text{int}} ,S^{\dagger}_{\mathbf{q}}\right] | 0 \rangle = E(\mathbf{q}) S^{\dagger}_{\mathbf{q}} | 0 \rangle  .
\end{eqnarray} 
We evaluate the  quartic operators from the interaction within the ground state,  $v^{\dagger}_{\mathbf{k}} | 0 \rangle=0$ and $ c_{\mathbf{k}} | 0 \rangle =0$, which is equivalent to an RPA-type diagrammatic treatment~\cite{Ando1997}. 
We obtain a non-linear equation for the exciton wave function $\varphi_{\mathbf{q}}(\mathbf{k})$ and its dispersion  $E(\mathbf{q})$.
Due to the fact that the interaction factorises, as $U \phi(\mathbf{k,q}) \phi^*(\mathbf{k',q})$, this can be cast as a simple implicit equation for   $E(\mathbf{q})$
\begin{eqnarray}
\label{DispersionEq}
1= -U \sum_{\mathbf{k}} \frac{|\phi(\mathbf{k,q})|^2 }{E(\mathbf{q})-\left[ E_{+}^{-}(\mathbf{k+q})-E_{-}^{+}(\mathbf{k})\right]} .
\end{eqnarray}

We show a representative exciton dispersion in the upper panel of Fig.\ref{Fig1b}. Generically, it has an almost degenerate ring-like manifold of energy minima at momenta $\mathbf{Q}$, see inset. This peculiar form of the dispersion originates from the special form of the TKI band structure where $\mathbf{Q}/2$ are the vectors pointing to the maxima (minima) of the bands. The precise value, $\Delta_{\text{Exc}}$, of the dispersion minimum depends on the strength of the residual interaction. 
Several experiments on SmB$_6$ have observed in-gap states reminiscent of our excitons. Inelastic neutron scattering (INS) measured weakly dispersing 'spin-excitons'~\cite{Fuhrman2015,Fuhrman2014}, with local minima close to the Brillouin zone boundaries similar to our calculation but with a sizeable gap.
Scanning tunneling spectroscopy~\cite{Ruan2014} and inelastic light scattering~\cite{Nyhus1997} find evidence for low energy collective modes. The notorious resistivity plateau at low temperatures, which was originally attributed to impurity bands~\cite{Allen1979}, has also been related to a kind of exciton-complexes which acquire charge by trapping electrons~\cite{Kikoin2000}. However, recent charge transport measurements on thin films \cite{Kim2013} have conclusively related the plateau to surface states which invalidates such a polaronic explanation.

\begin{figure}
\centering
\includegraphics[width=1.0\linewidth]{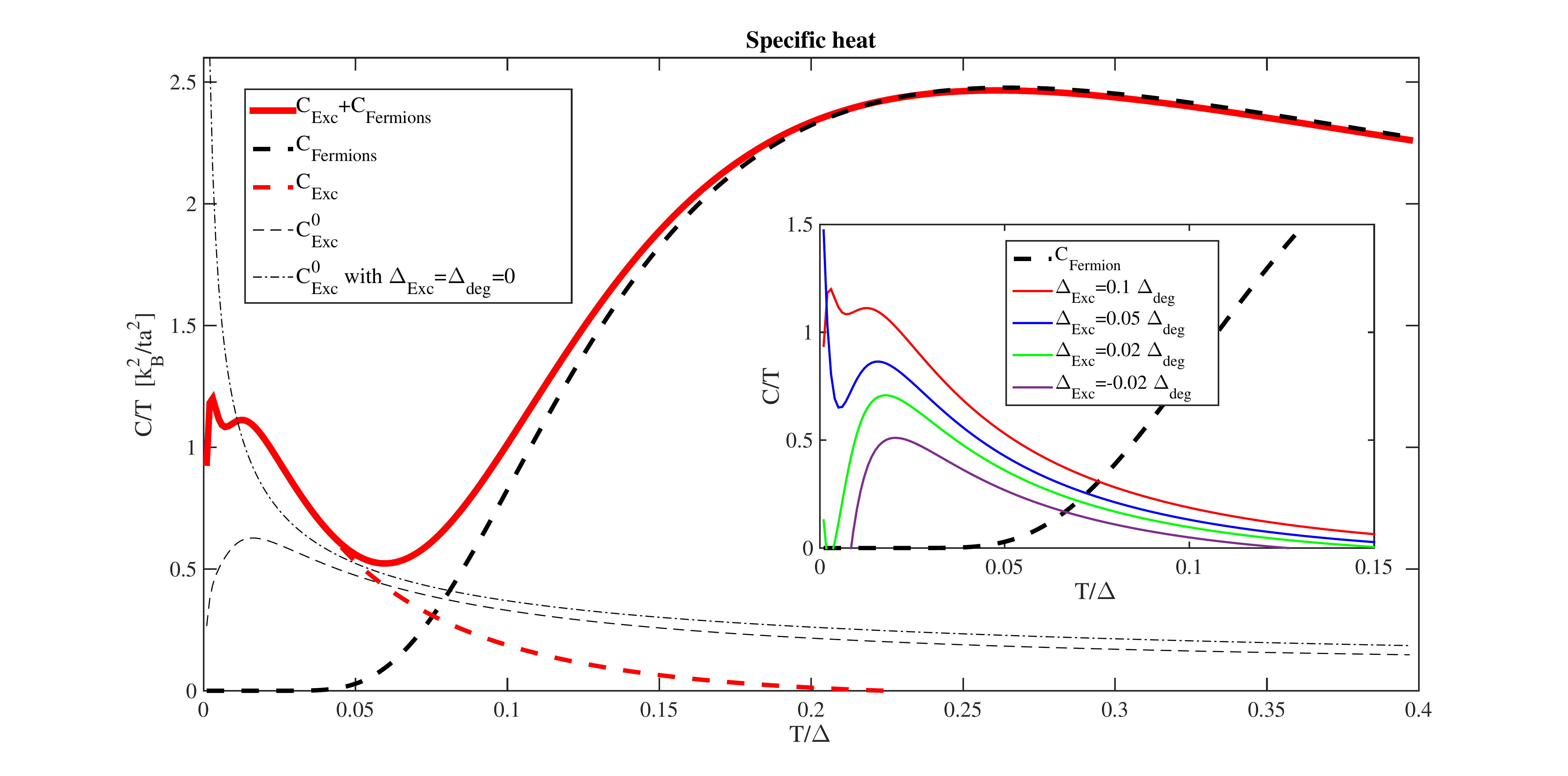}
\includegraphics[width=1.0\linewidth]{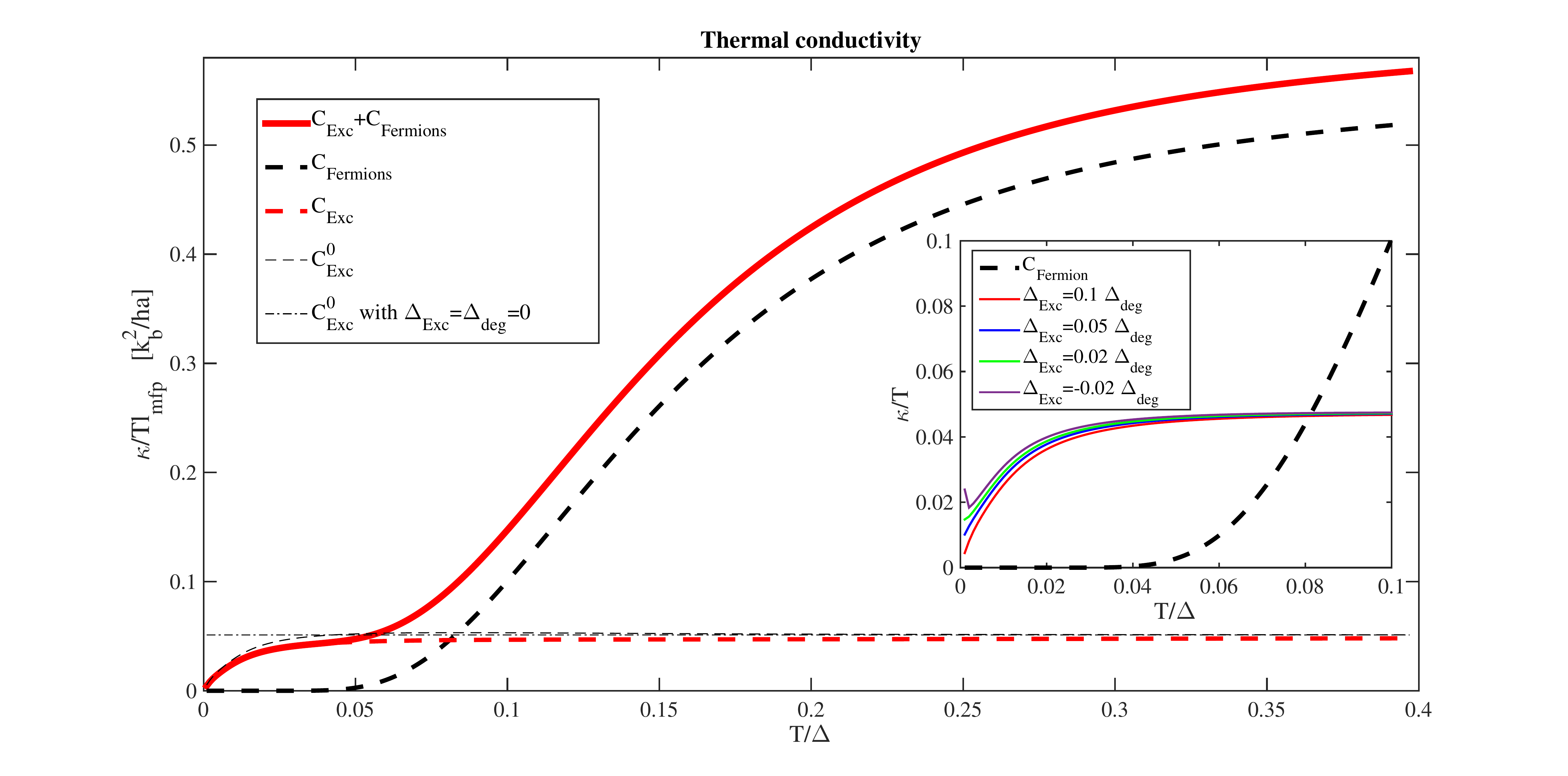}
\caption{{\bf Specific heat and thermal transport.} The upper panel shows the specific heat for $B=0$, plotted as $C/T$, as a function of scaled temperature $T/\Delta$ with the fermionic band gap $\Delta$. For comparison, we plot the behaviour of non-interacting excitons, $C_{\text{Exc}}^0$, with a dispersion with a perfect degenerate ring-like minimum (black dot dashed) or in the self-consistent lattice dispersion (black dashed). 
We use the same parameters as in the upper panel of Fig.\ref{Fig2} (with $\Delta_{\text{deg}}/\Delta=1/20$, $\Delta_{\text{Exc}}=\Delta_{\text{deg}}/10$, $g=0.004t$) and scaled the exciton contribution by a factor $1/4$.
The lower panel shows the thermal conductivity plotted as $\kappa/T$. }
\label{Fig2}
\end{figure}

We postulate that in SmB$_6$ the exciton binding energy is large such that the exciton gap $\Delta_{\text{Exc}}$ is small compared to the bandgap (and hence the activation gap from bulk resistivity) and sets a low temperature scale.
In the simplest approximation excitons can be treated as a non-interacting gas of bosons, with specific heat $C=\sum_{\mathbf{k}} E(\mathbf{k}) \frac{\partial n_{\mathbf{k}}}{\partial T}$
where  $ n_{\mathbf{k}} = \left[ e^{\frac{E(\mathbf{k})}{k_B T}}-1 \right]^{-1}$. The almost degenerate ring-like minimum of the excitons has strong consequences for the behaviour of experimental observables because it reduces the effective dimensionality to one dimension. For perfect degeneracy and $\Delta_{\text{Exc}}=0$ such that $E(\mathbf{q})\propto (|\mathbf{q}|-|\mathbf{Q}|)^2$, it is easy to show that a residual specific heat contribution $C/T\propto 1/\sqrt{T}$ appears independent of dimensionality!

 The apparent fine-tuning of $\Delta_{\text{Exc}}\approx 0$ is relaxed if one goes beyond the approximation of purely non-interacting bosons. First, lattice effects lift the degeneracy such that the divergence in specific heat would be removed below a scale $\Delta_{\text{deg}}$ which is the difference between the maximum and minimum energy on the ring. Second, Pauli blocking  suppresses the formation of excitons with similar momenta. Third, additional interactions between excitons alter the simple thermal occupation. We can model the influence of interactions, 
details of the self-consistent calculation of the thermal occupation, $n_{\mathbf{k}}$, are given in the supplementary material. The main effect is a suppression of the exciton number. The qualitative behaviour of the free boson approximation survive but the system is much less fine-tuned, e.g. even for a small negative gap the exciton density remains small. 

The resulting specific heat is shown in the upper panel of Fig.\ref{Fig2}. One clearly observes that excitons inside the band gap lead to an extra contribution at low temperature in contrast to a pure fermionic scenario (black dashed) which has exponential suppression at low $T$~\cite{Phelan2014}. 
The main panel shows the specific heat as calculated from the self-consistently determined exciton dispersion compared to the asymptotic behaviour of a non-interacting model with $\Delta_{\text{deg}}=0$ (thin dot dashed) and $\neq 0$ (thin dashed). 
In the inset of Fig.\ref{Fig2} we show $C/T$ calculated for different values of the exciton gap. We find a strong low temperature exciton contribution with an upturn very similar to experiments on SmB$_6$~\cite{Nickerson1971,Phelan2014}.

Charge neutral excitons cannot lead to charge transport, however, they can conduct heat. Within a semi-classical Boltzmann-like treatment we calculate their thermal conductivity as $\kappa=\sum_{\mathbf{k}} l_{\text{mfp}}  | v_{\mathbf{k}} | E(\mathbf{k}) \frac{\partial n_{\mathbf{k}}}{\partial T}$ with velocity $v_{\mathbf{k}}=\frac{\partial E(\mathbf{k})}{\partial \mathbf{k}}$ and the mean free path $l_{\text{mfp}}$ which we take to be constant  from impurity scattering. 
Excitons dominate  thermal conductivity at temperatures below the charge gap.
Their contribution originates again from the special form of the exciton dispersion, e.g. non-interacting bosons with a gapless dispersion with a perfectly degenerate ring-like minimum directly give $\kappa/T =$ const., thus, mimicking the behaviour of a metal. 
As shown in the lower panel of Fig.\ref{Fig2}, in our interacting lattice calculation this linear $\kappa$-term is  non-zero but strongly reduced for increasing $\Delta_{\text{Exc}}$, see inset.  
Since the precise value of the exciton gap is sensitive to microscopic details, there may be variations in the asymptotic low temperature behaviour between samples~\cite{Suchitra2016,Xu2016} arising from small changes in this gap.

{\it Magneto-excitons for $B>0$.}
Already a weak field has direct consequences in our scenario:  e.g. the lifting of degenerate exciton branches ($c^{\dagger}_- v_+$ and $c^{\dagger}_+ v_-$) from a Zeeman coupling, and a  reduction of the fermionic gap which leads to an increase of the thermal conductivity $\kappa$
and the specific heat anomaly~\cite{Flachbart2006}. 

 \begin{figure}
\centering
\includegraphics[width=1.0\linewidth]{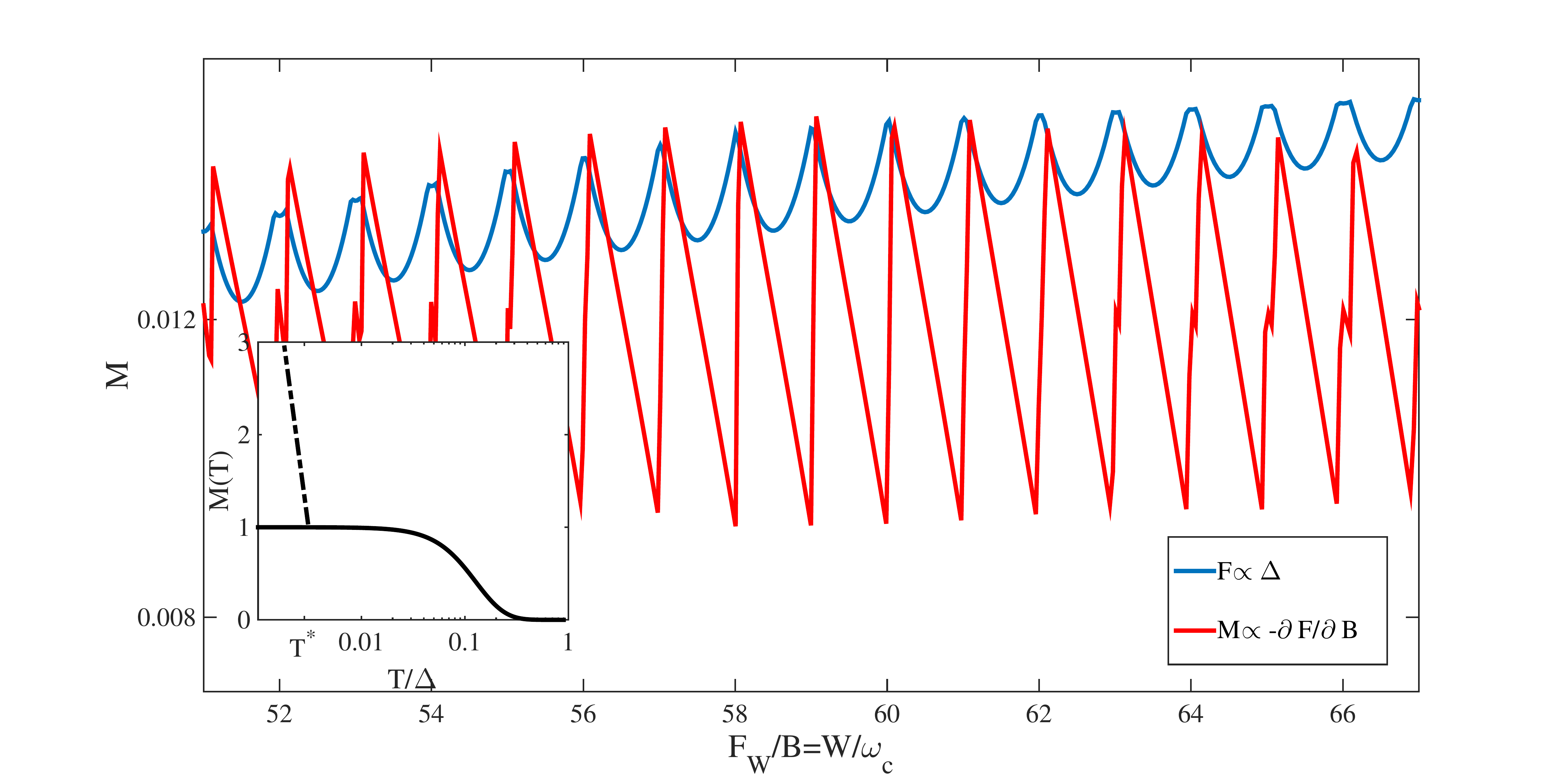}
\caption{{\bf Excitonic de Haas van Alphen effect.}  The energy of MExc, which is mainly given by the LL gap for weak MExc binding, changes periodically as a function of inverse magnetic field, see Eq.\ref{MOdisp1}. Therefore, the free energy of the exciton system $F$ (blue) varies periodically together with the corresponding magnetisation $M$ (red) (same parameters as before and $\gamma/W=0.025$). The schematic temperature dependence is shown in the inset.}
\label{MOfromME}
\end{figure}  
But could excitons also lead to a de Haas van Alphen effect (dHvAe)?
Of course, the MExc itself is charge neutral and its center of mass does not undergo cyclotron motion due to the absence of a Lorentz force~\cite{Erten2016}. However, they are built of particle and hole bands that form discrete Landau levels (LLs) in a magnetic field which in turn can lead to QO even for band insulators~\cite{Knolle2015,Wang2016}. Hence, the energy of the MExc varies periodically as a function of $1/B$ through the variation of the energy of its constituents. 
This ultimately simple mechanics (abbreviated as EdHvAe) leads to an oscillatory behaviour of the total energy as a function of $1/B$ and hence the magnetisation. 

The temperature dependence of the EdHvAe can be  estimated qualitatively. 
We have shown recently~\cite{Knolle2015} that narrow gap insulators can exhibit an {\it anomalous} de Haas van Alphen effect (AdHvAe) if the cyclotron frequency $\hbar \omega_c$ is comparable to the activation gap of the insulating state, see also Ref.\cite{Wang2016}. SmB$_6$ has an activation gap of about $\approx 10$ meV~\cite{Frantzeskakis2013,Fuhrman2015} and a small effective mass of the itinerant d-band \cite{Frantzeskakis2013,Lu2013}, $m_d=0.44 m_e$, leading to $\hbar \omega_c \approx 4 \mbox{meV}$  for 15 Tesla. Hence, the $B=0$ gap is slightly bigger than the cyclotron energy, however, in a TKI the gap is reduced in a magnetic field~\cite{Wang2016}. Overall, we expect that the amplitude of oscillations from the AdHvAe is small but otherwise follows a Lifshitz-Kosevich temperature dependence~\cite{Knolle2015,Wang2016} at higher temperatures. 
Since the EdHvAe also originates from the periodic variation of gapped LL branches we expect a temperature dependence similar to that of the AdHvAe but with an enhanced intensity.

It is tempting to speculate about the low temperature contribution of MExc which is complicated by the appearance of several low energy scales, e.g $\Delta_{\text{deg}}, \Delta_{\text{Exc}}$. 
For $\Delta_{\text{Exc}}<0$ a condensate should form at a temperature $T^*$ below which the density of MExc quickly grows leading to a steeply increasing QO amplitude. A schematic plot of such a temperature dependence is shown in the inset of Fig.\ref{MOfromME}. For $\Delta_{\text{Exc}}>0$ we would expect a thermally activated number of excitons (and contribution to QO) for $T<\Delta_{\text{Exc}}$. However, when projecting the f-level repulsion onto the TKI bands we have only concentrated on those terms which lead to direct exciton binding, but there exist residual terms which create or destroy pairs of $(+-)$  and $(-+)$ excitons. On the one hand, these exciton-non-conserving terms smoothen out the condensation transition to a mere crossover. On the other hand, they lead to a correction to the ground state energy from virtual exciton-pair fluctuations which entails a contribution to the QO amplitude even in the absence of thermally excited excitons for $\Delta_{\text{Exc}}>0$ and $T\to0$.

We can corroborate our qualitative discussion of the EdHvAe by a microscopic calculation in the strong field limit.
We only sketch the main steps  of our MExc calculation from the microscopic model, Eq.\ref{Hamiltonian}, details are relegated to the supplementary material. First, we calculate the energy spectrum and wave functions of $H_0$ in a strong magnetic field. Second, we can project the f-level interaction $H_{\text{int}}$ onto the LLs. Third, we derive a Bethe-Salpeter equation for MExc.
Similar to the $B=0$ case, the interaction only binds particle-hole pairs from opposite sectors $\nu$. For a given magnetic field there is a LL index $N_-$ ($N_+$) such that the energy of the corresponding valence (conduction) LL branch $E^{\nu}_- (N_-)$ ($E^{\bar \nu}_+(N_+)$) is maximal (minimal). We study the formation of excitons between those extremal levels and obtain a closed expression for the ME dispersion
\begin{widetext}
\begin{eqnarray}
\label{MOdisp1}
E(\mathbf{k}) & = & \Delta(B) - U \left[\cos \frac{\beta^-_{N_+}}{2} \right]^2 \left[\sin \frac{\beta^+_{N_-}}{2} \right]^2 \frac{(-1)^{N_- +N_+-1}}{2 \pi} e^{-\frac{k^2}{2}} L_{N_+-1}^{N_- -N_++1} \left( \frac{k^2}{2}\right) L_{N_-}^{N_+-1-N_-} \left( \frac{k^2}{2}\right)
\end{eqnarray}
\end{widetext}
with the LL gap $\Delta (B)=E_+^-(N_+)-E_-^+(N_-)$ and the angles given by $\tan \beta_n^{\nu} = \frac{ \hbar \gamma}{2\sqrt{2} l_B} \frac{ \sqrt{n}}{ \hbar  \omega_c \left ( n - \nu \frac{1}{2} \right )+\alpha  \hbar  \omega_c \left ( n + \nu \frac{1}{2} \right ) -\lambda }$ ($ l_B=\sqrt{\frac{c\hbar}{eB}}, \omega_c=\frac{eB}{m_d c}$).

In the regime $E_B<\hbar \omega_c<\Delta(B)$ the change of the exciton dispersion $E(\mathbf{k})$ as a function of field will be mainly determined by a change in the LL gap $\Delta(B)$, see Eq.\ref{MOdisp1}. Then, for a given temperature the free energy and the magnetisation are directly related to the periodic variation of the gap $M=-\frac{\partial F}{\partial B} \propto -\frac{\partial \Delta}{\partial B}$. In Fig.\ref{MOfromME} we plot this oscillatory behaviour as a function of $W/\hbar \omega_c = F_W/B$. There is indeed an exciton contribution to the magnetisation whose period $1/F_W$ corresponds to a dHvAe frequency set by the area of the intersection of the unhybridised d- and f-bands similar to experimental findings in SmB$_6$~\cite{Tan2015}.

{\it Discussion.}
In our excitonic scenario we expect that applying a magnetic field should increase $C,\kappa$ and $T^*$ due to an increase in exciton binding energy.  To settle the ongoing debate whether the experimentally observed QOs are coming from gapless surface states \cite{Fisk2014,Erten2016,Thomson2016}, or from the bulk \cite{Tan2015,Knolle2015,Wang2016} one should directly compare the field dependence of charge transport [Shubnikov de Haas effect (SdHe)], and thermodynamic quantities (dHvAe). If both were only due to gapless surface states the behaviours of SdHe and dHvAe should be similar to those in a metal. However, if one is a bulk signal and the other from the surface we expect that both differ strongly.

There are strong predictions to test our scenario: First, despite the absence of a bulk SdHe there should be QO in the {\it thermal} conductivity which would also allow a clear separation of the exciton contribution from phonons. Second, we expect a low energy mode in INS experiments.

In conclusion, we have shown that TKIs are very susceptible to the formation of excitons because of their correlated origin. 
These charge neutral but spinful quasiparticles have a dispersion with a ring-like minimum which gives rise to a specific heat contribution and residual thermal conductivity mimicking the low temperature behaviour of a metallic system in agreement with long standing observations in SmB$_6$. 
We showed that the magneto-exciton energy varies periodically as a function of inverse magnetic field leading to an unexpected dHvAe in an insulator without a Fermi surface. 

\section{Acknowledgements}
We  thank G. Lonzarich and S. Sebastian for helpful discussions and for sharing their experimental data. 
 The work is supported by a Fellowship within the Postdoc-Program of the German Academic Exchange Service (DAAD) and by the EPSRC Grant No. EP/J017639/1.

\vspace{2.5cm}
\begin{widetext}
\section{Supplementary Material}	
\subsection{Thermal occupation of excitons}
Here, we outline our calculation of the thermal occupation of excitons. As stated in the main text we expect that excitons are bosonic quasiparticles with interactions  dictated by the additional terms of $H_{int}$ and Pauli blocking, which limits the number of excitons in the same momentum state. As a full treatment of the interacting theory is beyond the scope of our work we estimate the important qualitative effects. 

There are two natural length scales: the exciton size $\xi$ related to the exciton binding energy $E_B$ and roughly given by $E_B=\frac{\hbar^2}{2m_{c,v}} \frac{1}{\xi^2}$, and the thermal wave length $\lambda_T$ of excitons from $k_B T= \frac{\hbar^2}{2m_{\text{Exc}}} \frac{1}{\lambda_T^2} $.
We expect that $E_B$  is roughly of the size of the fermionic gap which is much larger than the low temperature regime we are interested in $k_B T \ll \Delta \approx E_B$. Therefore $\xi \ll \lambda_T$ and only excitons in a small patch of momentum space of size $A_\varphi \approx \frac{1}{\xi^2}$ effectively interact.

A natural energy functional for excitons with interactions short ranged in momentum space is
 \begin{eqnarray}
 E_{\text{tot}} =  \sum_{\mathbf{k}} n_{\mathbf{k}} E(\mathbf{k}) + \label{EnergyFunctional1}+ \tilde g \sum_{\mathbf{k},\mathbf{k'}} e^{-\frac{\left(\mathbf{k} -\mathbf{k'}\right)^2}{\xi^2}}n_{\mathbf{k}} n_{\mathbf{k'}},
\end{eqnarray}
which we approximate in the low temperature regime by a sum  $E_{\text{tot}}=\sum_{\varphi} E_{\text{tot}}^\varphi$ over small patches (labeled by $\varphi$) of size $A_\varphi$ making up the total BZ  in which the interaction is taken to be constant
\begin{eqnarray}
 E_{\text{tot}}^\varphi =   \sum_{\mathbf{k} \in \varphi} n_{\mathbf{k}} E(\mathbf{k}) + \frac{g}{2} \left( \sum_{\mathbf{k} \in \varphi} n_{\mathbf{k}} \right)^2 . 
 \label{EnFunct2}
\end{eqnarray}

Introducing the mean number of excitons in each patch, $N^\varphi= \sum_{\mathbf{k} \in \varphi} \langle n_{\mathbf{k}} \rangle$, this can be further simplified by neglecting  number fluctuations 
\begin{eqnarray}
 E_{\text{tot}}^\varphi =   \sum_{\mathbf{k} \in \varphi} n_{\mathbf{k}} \left[ E(\mathbf{k}) +g N^\varphi \right] - \frac{1}{2} g \left( N^\varphi \right)^2  + \xcancel{\frac{g}{2} \left[ \sum_{\mathbf{k} \in \varphi} n_{\mathbf{k}} - N^\varphi \right]^2 }. 
 \label{EnFunct3}
\end{eqnarray}
This mean-field functional is linear in the occupation numbers such that we directly obtain a self-consistency equation for the exciton number in each patch
\begin{eqnarray}
N^\varphi =   \sum_{\mathbf{k} \in \varphi} \frac{1}{e^{\frac{E(\mathbf{k})+g N^\varphi}{k_B T}}-1} .
 \label{SelfConistsency1}
\end{eqnarray}
We solve this numerically for a given exciton dispersion to directly obtain the thermal occupation (here $N(\mathbf{k})=N^\varphi$ for all $\mathbf{k}\in\varphi$ )
\begin{eqnarray}
n_{\mathbf{k}} =   \frac{1}{e^{\frac{E(\mathbf{k})+g N(\mathbf{k})}{k_B T}}-1} 
 \label{SelfConistsency2}
\end{eqnarray}
which enters the calculation of the specific heat and thermal conductivity, see main text.
In practice, we divide the BZ into $p$ patches each of which is a wedge with an angle $\phi=\frac{2 \pi}{p}$. We have confirmed that the results do not change qualitatively for large enough $p$ and show calculations in the main text for $p=24$.

\subsection{Magneto exciton calculation}

Here, we give details of the MExc calculation. We work in a continuum approximation, e.g. $\epsilon_{\mathbf{k}} \approx \frac{k^2}{2m_d}$ and $\sin {k_{\eta}} \approx k_{\eta}$ which allows an exact calculation of the LL spectrum. We introduce the orbital magnetic field in the $z$-direction via the vector potential $\mathbf{B}=\nabla \times \mathbf{A}$ with $\mathbf{A}=(0,Bx,0)$ in the Landau gauge. Note, the Zeeman energy is expected to be negligibly small in SmB$_6$~\cite{Tan2015}.
The vector potential ${\bf A}$ is minimally coupled to the crystal momentum such that ${\bf \Pi}= {\bf k}-\frac{e}{c} {\bf A}$ is the gauge invariant momentum. The quadratic d- and f- level dispersions can be written in terms of the standard raising and lowering operators $a = \frac{l_B}{\sqrt{2}\hbar} \left (\Pi_x - \text{i} \Pi_y \right )$  and $  a^{\dagger}= \frac{l_B}{\sqrt{2}\hbar} \left (\Pi_x + \text{i} \Pi_y \right ) $ with $  \left [ a,a^{\dagger} \right ]=1 $ and $ \left[ \Pi_x,\Pi_y \right ]=-\text{i}\frac{\hbar^2}{l_B^2}$.
We solve the eigenvalue equation $ H_0 |\Psi \rangle = E_n |\Psi \rangle$ with the Ansatz $| \Psi \rangle= \left (u_n |n-1\rangle, v_n |n\rangle, x_n |n-1\rangle, y_n |n\rangle    \right )^T$ and the standard harmonic oscillator states $| n \rangle$ ($ l_B=\sqrt{\frac{c\hbar}{eB}}, \omega_c=\frac{eB}{m_d c}$)
\begin{eqnarray}
\label{HamiltonianContinuumNnumers}
\!\! \! \! \! \! \!
\begin{pmatrix}
 \hbar  \omega_c \left ( n - \frac{1}{2} \right ) +\frac{k_z^2}{2m}& 0 & \frac{\gamma}{2} k_z &  \frac{\sqrt{2} \hbar \gamma}{2 l_B} \sqrt{n} \\
0 &   \hbar  \omega_c \left ( n + \frac{1}{2} \right ) +\frac{k_z^2}{2m}  &  \frac{\sqrt{2} \hbar \gamma}{2 l_B} \sqrt{n} & -\frac{\gamma}{2} k_z\\
 \frac{\gamma}{2} k_z &  \frac{\sqrt{2} \hbar \gamma}{2 l_B} \sqrt{n} & -\alpha  \hbar  \omega_c \left ( n- \frac{1}{2} \right ) - \alpha \frac{k_z^2}{2m} +\lambda & 0\\
 \frac{\sqrt{2} \hbar \gamma}{2 l_B} \sqrt{n} & -\frac{\gamma}{2} k_z & 0 & -\alpha  \hbar  \omega_c \left ( n + \frac{1}{2} \right ) - \alpha \frac{k_z^2}{2m} +\lambda
 \end{pmatrix}
 \! \! \!
 \begin{pmatrix}
 u_n \\ v_n\\ x_n\\ y_n
 \end{pmatrix} 
\!\! = \!  E_n \! 
  \begin{pmatrix}
 u_n \\ v_n\\ x_n\\ y_n
 \end{pmatrix}   .
 \end{eqnarray}
Energy levels can be obtained for arbitrary $k_z$ but even for the case of $B=0$ the effect of the third dimension is negligible and an orbital magnetic field makes the system even more two dimensional; for example it is only the extremal orbit at $k_z=0$ which determines the main oscillation period of the standard dHvAe~\cite{Shoenberg1984}. 
For $k_z=0$ the calculation of the spectrum decouples  again into two independent eigenvalue problems ($d_{\uparrow} f_{\downarrow}$ and $d_{\downarrow} f_{\uparrow}$). Care needs to be taken for the $n=0$ energies because of $a|0\rangle =0$ which leads to $u_{n=0}=0$ and $x_{n=0}=0$. For $n=0$ the two non-degenerate energies are
$E^{\uparrow \downarrow} (n=0) = -\alpha \frac{\hbar \omega_c}{2} +(1+\alpha) W $ and $E^{ \downarrow \uparrow} (n=0) =  \frac{\hbar \omega_c}{2}$. 
For $n>0$ there are four energies
\begin{eqnarray}
\label{LLbranches}
E_{\pm}^{\nu}(n)\! = \! \frac{1}{2} \! \left(\! \hbar \omega_c\! \left( \! n \! - \! \frac{\nu}{2} \! \right) \!-\! \alpha \hbar \omega_c \! \left(\! n\! +\! \frac{\nu}{2} \! \right)\! +\! (\!1+\!\alpha \! )\! W \!\pm \! \sqrt{\! \left [\! \hbar \omega_c\! \left (\! n\! -\! \frac{\nu}{2}\! \right )\! +\!\alpha \hbar\omega_c \!\left (\! n\! +\frac{\nu}{2}\! \right )\! -\! (1\!+\!\alpha)\! W \!\right ]^2 \!+\! \left [\! \frac{\sqrt{2}\! \hbar \gamma}{l_B} \!\right ]^2 \!n}\!  \right ).
\end{eqnarray}
However, in contrast to $B=0$ they are all non-degenerate because the momentum dependent hybridisation mixes different LLs. This property, which is intimately linked to the topological nature of the band structure, also leads to a magnetic field and LL index dependent gap between the valence and conduction branches~\cite{Wang2016} which would be absent in a momentum independent hybridisation, e.g. for a simple KI.
We have a 'conduction' and a 'valence' LL branch for each  $\nu$ such that $H_0 = \sum_{n,p,\nu} \left[ E_{+}^{\nu}(n) c_{n,p,\nu}^{\dagger} c_{n,p,\nu} +  E_{-}^{\nu}(n) v_{n,p,\nu}^{\dagger} v_{n,p,\nu} \right] $. 

Next, we project the Hubbard interaction onto the LLs. In second quantised form the field operators, e.g. for the f-electrons, are written as $\Psi_f^{\dagger} (\mathbf{r})= \sum_{n,p} \psi_{n,p} (\mathbf{r})f_{n,p}^{\dagger}$ with the normalised harmonic oscillator wave functions~\cite{Bychkov1983}
\begin{eqnarray}
\psi_{n,p} (\mathbf{x,y}) & = & \frac{1}{\sqrt{L}} e^{\text{i} p y} \left[ \pi 2^{2n} (n!)^2\right]^{-\frac{1}{4}} e^{-\frac{1}{2} (x+p)^2} H_n(x+p)
\end{eqnarray}
with the Hermite polynomials $H_n(x)$. Note, all coordinates are measured in units of magnetic length $l_B=\sqrt{\frac{c\hbar}{eB}}$ such that $x,y \to l_B x,l_By$ and $p\to p/l_B$. 

From Eq.\ref{HamiltonianContinuumNnumers} we not only get the spectrum but directly obtain the wave functions for the different LL branches for $n>0$
\begin{eqnarray}
c_{n,p,+} & = & \cos \frac{\beta_n^{+}}{2} d_{n-1,p,\uparrow} + \sin \frac{\beta_n^{+}}{2} f_{n,p,\downarrow} \\ \nonumber
v_{n,p,+} & = & -\sin \frac{\beta_n^{+}}{2} d_{n-1,p,\uparrow} + \cos \frac{\beta_n^{+}}{2} f_{n,p,\downarrow} \\ \nonumber
c_{n,p,-} & = & \cos \frac{\beta_n^{-}}{2} d_{n,p,\uparrow} + \sin \frac{\beta_n^{+}}{2} f_{n-1,p,\downarrow} \\ \nonumber
v_{n,p,-} & = & -\sin \frac{\beta_n^{-}}{2} d_{n,p,\uparrow} + \cos \frac{\beta_n^{+}}{2} f_{n-1,p,\downarrow} 
\end{eqnarray}
with the angles given by $\tan \beta_n^{\nu} = \frac{ \hbar \gamma}{2\sqrt{2} l_B} \frac{ \sqrt{n}}{ \hbar  \omega_c \left ( n - \nu \frac{1}{2} \right )+\alpha  \hbar  \omega_c \left ( n + \nu \frac{1}{2} \right ) -\lambda }$.

Next we project $H_{\text{int}}$ on to the diagonal LL basis~\cite{Lerner1981,Bychkov1983,Kallin1984,MacDonald1985,Bychkov2008,Toeke2011} and a lengthy calculation gives (again we only show the contribution which leads to MExc binding)
\begin{eqnarray}
H_{\text{int}}  =  U \sum_{n_1...n_4,p_1,p_2,q_x,q_y} e^{\text{i}q_x (p_1-p_2-q_y)} && \left[ J^{\alpha}_{n_4,n_1} (\mathbf{q}) J^{\beta}_{n_3,n_2} (-\mathbf{q})
c_{n_1+1,p_1,-}^{\dagger} v_{n_2,p_2,+}^{\dagger}v_{n_3,p_2+q_y,+} c_{n_4+1,p_1-q_y,-} \right.+ \\ \nonumber
 &&  \left. J^{\gamma}_{n_4,n_1} (\mathbf{q}) J^{\delta}_{n_3,n_2} (-\mathbf{q})
v_{n_1+1,p_1,-}^{\dagger} c_{n_2,p_2,+}^{\dagger}c_{n_3,p_2+q_y,+} v_{n_4+1,p_1-q_y,-}  \right]
\end{eqnarray}
with 
\begin{eqnarray}
J^{\alpha}_{n_4,n_1} (\mathbf{q}) & = & \cos \frac{\beta^-_{n_1+1}}{2}  \cos \frac{\beta^-_{n_4+1}}{2} J_{n_4,n_1} (\mathbf{q}) \\ \nonumber
J^{\beta}_{n_3,n_2} (\mathbf{q}) & = & \sin \frac{\beta^+_{n_2}}{2}  \sin \frac{\beta^+_{n_3}}{2} J_{n_3,n_3} (\mathbf{q}) \\ \nonumber
J^{\gamma}_{n_4,n_1} (\mathbf{q}) & = & \sin \frac{\beta^-_{n_1+1}}{2}  \sin \frac{\beta^-_{n_4+1}}{2} J_{n_4,n_4} (\mathbf{q}) \\ \nonumber
J^{\delta}_{n_3,n_2} (\mathbf{q}) & = & \cos \frac{\beta^+_{n_2}}{2}  \cos \frac{\beta^+_{n_3}}{2} J_{n_3,n_2} (\mathbf{q})  \\
\text{and} \  \  J_{m,n} (q_x,q_y) & = & \sqrt{\frac{n!}{m!}} e^{\frac{q^2}{4}} \left( \frac{q_y-\text{i}q_x}{\sqrt{2}}\right)^{m-n} L_n^{m-n} \left( \frac{q^2}{2}\right)
\end{eqnarray}
and the generalized Laguerre Polynomials $L_n^{m-n} \left( x\right)$~\cite{Bychkov1983}.

Again, the interaction only binds particle-hole pairs from opposite blocks $\nu$. For a given field $B$ there is a LL index $N_-$ ($N_+$) such that the energy of the corresponding valence (conduction) LL branch $E^{\nu}_- (N_-)$ ($E^{\bar \nu}_+(N_+)$) is maximal (minimal). We study the formation of excitons between those extremal levels which is equivalent to the approximation~\cite{Bychkov1983,Kallin1984,MacDonald1985,Bychkov2008,Toeke2011} that the ME binding energy is smaller than the inter LL spacing.
The MExc creation operator is~\cite{Bychkov1983,Bychkov2008}
\begin{eqnarray}
\label{MagnetoExcOperator}
S_{\mathbf{k}} & = & \sum_p e^{\text{i} k_x(p+\frac{k_y}{2})}  \varphi_{\mathbf{k}} (p) c_{N_+,p,-}^{\dagger} v_{N_-,p+k_y,+}.
\end{eqnarray}

From the Bethe-Salpeter equation, see Eq.\ref{ExcitonEqOfMotion}, we obtain the MExc dispersion. Under the assumption that the projected interaction primarily mixes particle-hole excitations with LL indices $N_+$ and $N_-$ (which is rigorous in the limit $\left[ E_{+}(N_+)-E_{-}(N_-) \right] \ll \left[E_{+}(n)-E_{-}(m) \right]$) we get
\begin{eqnarray}
E(\mathbf{k}) & = & E_+^-(N_+)-E_-^+(N_-) - U \sum_{q_x,q_y} e^{-\text{i} (q_x k_y+ q_y k_x)} J^{\alpha}_{N_++1,N_++1} (\mathbf{q}) J^{\beta}_{N_-,N_-} (-\mathbf{q}).
\end{eqnarray}
There is an additional constant energy shift $C=U \sum_{n_2} J^{\alpha}_{N_++1,N_++1} (0) J^{\beta}_{n_2,n_2} (0) $ which is formally divergent because of the continuum approximation of the inverted band similar to the case of graphene~\cite{Bychkov2008}. However, this constant is canceled by a diverging contribution of the opposite sign from the ground state energy. 

Finally, we obtain a closed expression for the dispersion via integral relations of the generalised Laguerre polynomials~\cite{Koelbig1996}
\begin{eqnarray}
\label{MOdisp}
E(\mathbf{k}) & = & E_+^-(N_+)-E_-^+(N_-) - U \left[\cos \frac{\beta^-_{N_+}}{2} \right]^2 \left[\sin \frac{\beta^+_{N_-}}{2} \right]^2 I(k)  \\ \nonumber
I(k) & = & \int_0^{\infty} \frac{\text{d}q}{2\pi} J_0(kq) q e^{-\frac{q^2}{2}} L_{N_+-1}\left( \frac{q^2}{2} \right) L_{N_-}\left( \frac{q^2}{2} \right) = \frac{(-1)^{N_- +N_+-1}}{2 \pi} e^{-\frac{k^2}{2}} L_{N_+-1}^{N_- -N_++1} \left( \frac{k^2}{2}\right) L_{N_-}^{N_+-1-N_-} \left( \frac{k^2}{2}\right).
\end{eqnarray}

The MExc dispersion is presented for two different values of $N_+$ in Fig.\ref{Fig4}. Note, different $N_+$ correspond to different magnetic fields and for the relevant regime $\hbar \omega_c \lesssim \gamma \ll W$ in SmB$_6$ we expect exciton pairing between high LLs. There are only two possibilities for the LL indices defining the minimal gap between the upper and lower branch $N_-=N_+$ (dashed) or $N_-=N_++1$ (solid). The corresponding ME dispersions have a minimum at finite or zero momentum and there are additional characteristic oscillations whose period scales with the `Fermi wavelength' $1/Q \propto \lambda_F \propto l_B/\sqrt{N_+}$. 
In addition, we find from an expansion of the Laguerre polynomials that $E(k)$ is quadratic in momentum around the minimum and the mass scales linearly with $N_+$ for small $k$.
Beyond our  approximation MEs are expected to be mixtures of more than one valence and conduction LL index which generically will lead to small hybridisation gaps between the crossings of the solid and dashed lines.

Overall, we expect a critical magnetic field $B_c$ where the exciton dispersion changes its qualitative behaviour transforming a ring-like dispersion minimum into a single minimum at high fields $B>B_c$.

\begin{figure}
\centering
\includegraphics[width=0.5\linewidth]{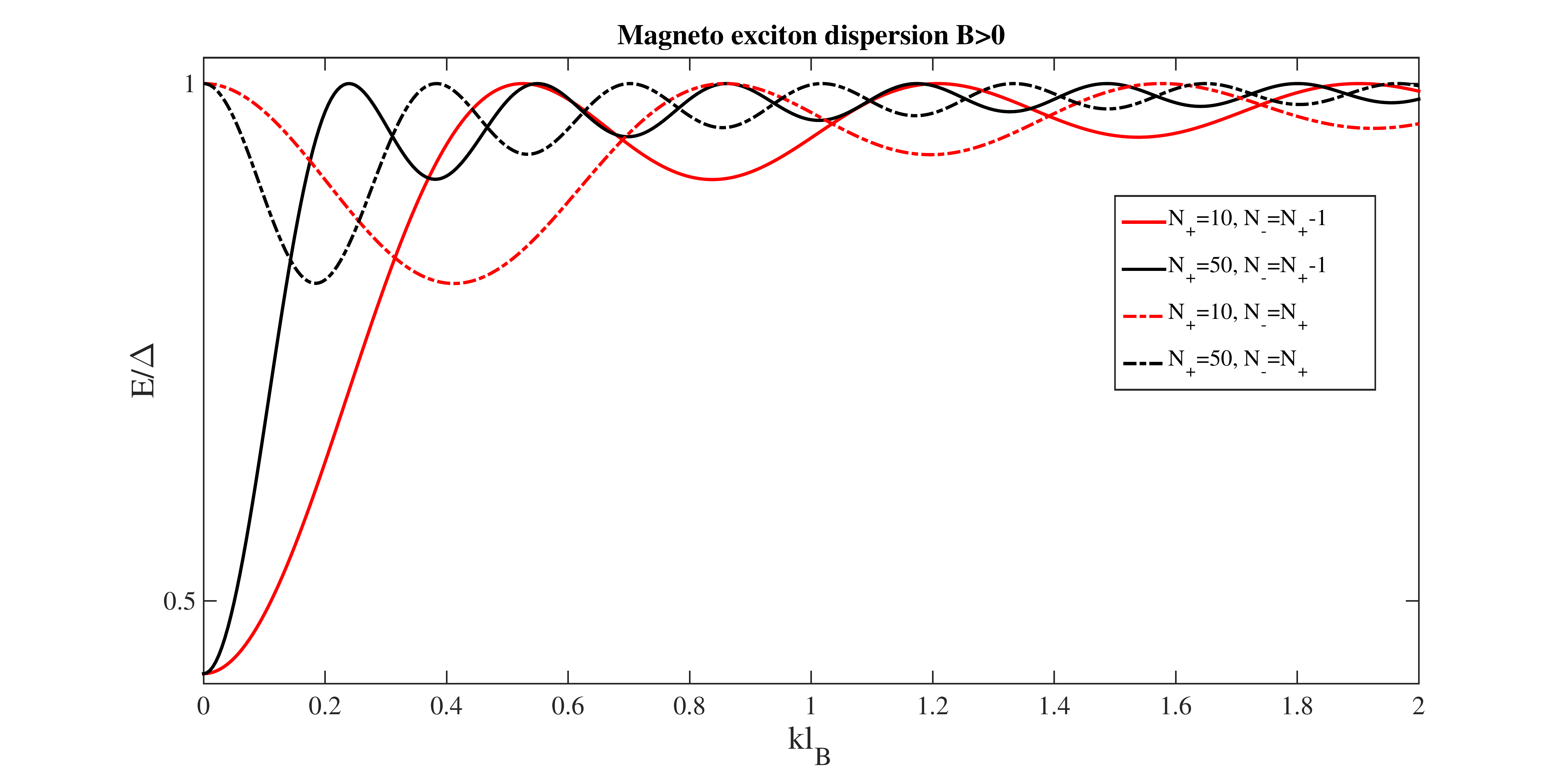}
\caption{{\bf Magneto-exciton dispersion.} The dispersion for a strong orbital magnetic field $B$, see text for discussion.
}
\label{Fig4}
\end{figure}

\end{widetext}

\end{document}